\documentclass[showpacs,aps,preprint]{revtex4}
\usepackage{graphicx}
\usepackage{amsmath,amssymb}

\def\gtsim{\mathrel{\hbox{\raise0.2ex
  \hbox{$>$}\kern-0.75em\raise-0.9ex\hbox{$\sim$}}}}
\def\ltsim{\mathrel{\hbox{\raise0.2ex
  \hbox{$<$}\kern-0.75em\raise-0.9ex\hbox{$\sim$}}}}

\newcommand{\half}{{\frac{1}{2}}}
\newcommand{\Jonem}[0]{{\cal J}^{(1)}}
\newcommand{\Jtwom}[0]{{\cal J}^{(2)}}

\begin{document}

\title{
Importance of multicranked configuration mixing for
angular-momentum-projection calculations:\\
Study of superdeformed rotational bands in $^{152}$Dy and $^{194}$Hg
}
\author{Masaki Ushitani, Shingo Tagami, and Yoshifumi R. Shimizu}
\affiliation{Department of Physics, Graduate School of Science,
Kyushu University, Fukuoka 819-0395, Japan}


\begin{abstract}
Recently we have investigated an effective method of multicranked
configuration-mixing for angular-momentum-projection calculation,
where several cranked mean-field states are coupled after projection:
The basic idea was originally proposed by Peierls and Thouless
more than fifty years ago.
With this method a good description of the rotational band has been
achieved in a fully microscopic manner.
In the present work, we apply the method to the high-spin superdeformed band,
for which long rotational sequence is observed,
and study how the good description is obtained for
the rotational spectrum as well as the $\Jonem$ and $\Jtwom$ moments of inertia
as functions of angular momentum.
The Gogny D1S force is employed as an effective interaction,
and the yrast superdeformed bands in $^{152}$Dy and $^{194}$Hg
are taken as typical examples
in the $A\approx 150$ and $A\approx 190$ regions, respectively.
The effect of pairing correlations is examined by
the variation after particle-number projection approach
to understand the different behaviors of $\Jtwom$ moments of inertia
observed in these two nuclei.
The particle-number projection
on top of the angular-momentum projection has been 
performed for the first time with the multicranked configuration-mixing.

\end{abstract}

\pacs{21.10.Re, 21.60.Ev, 23.20.Lv}

\maketitle

\section{Introduction}
\label{sec:intro}

Collective motion in atomic nuclei has been an interesting subject
in nuclear structure physics~\cite{BM75,RS80}.
The rotational motion is a typical collective motion and exhibits
many interesting phenomena especially at high-spin states,
see e.g. Refs.~\cite{VDS83,GHH86,Fra01,WP01,SW05,Fra18}.
Recently, we have developed a theoretical framework
for describing the high-spin rotational band in a fully microscopic manner
by employing the technique of angular-momentum projection
from the selfconsistent mean-field states~\cite{STS15}.
We called it ``angular-momentum-projected multicranked configuration-mixing'',
where several selfconsistently cranked mean-field states are
quantum-mechanically coupled after projection.
It has been shown that good agreement of the spectra and the
kinematic moments of inertia, $\Jonem$, is obtained
for the ground-state rotational bands in rare earth nuclei
without any adjustable parameters~\cite{STS16}.
In contrast to the projected shell-model approach,
where the angular-momentum projection method is successfully applied,
see e.g. Refs.~\cite{HS95,Sun16,SBD16},
the number of mean-field states coupled after projection
is relatively small in our approach,
where they are obtained selfconsistently
within the cranking procedure.

Generally, calculation of the mass parameter is crucial
for the appropriate description of nuclear collective motion.
It has been well-known that the generator coordinate method (GCM)
with only collective coordinates does not give proper mass parameter,
see e.g. Sec.\,\*11.4.5 of Ref.~\cite{RS80} for the instructive argument
especially for the center-of-mass motion.
The angular-momentum projection procedure is a special case of the GCM
for the collective rotational motion
with the Euler angles as collective coordinates.
In order to obtain the proper mass,
Peierls and Thouless proposed to superpose
not only wave functions with different coordinates but also those with
different velocities (or momenta)~\cite{PT62},
which incorporates time-odd components into the wave function;
the correct total mass appears for the center-of-mass motion
as a result of the boost because of the Galilean invariance.
In fact the time-odd components are generally important for mass parameters,
and we have investigated a procedure to take them into account,
which we call ``infinitesimal cranking'',
and successfully applied it to the collective $\gamma$-vibration
in Ref.~\cite{TS16}.

The basic idea of the framework of
multicranked configuration-mixing~\cite{STS15}
is nothing else but the one proposed by
Peierls and Thouless for the rotational motion~\cite{PT62},
i.e., the total wave function is calculated as
\begin{equation}
 |\Psi^I_{M,\alpha}\rangle =
 \int d\omega_{\rm rot}\sum_{K} g^I_{K,\alpha}(\omega_{\rm rot})\,
 \hat P^I_{MK}|\Phi(\omega_{\rm rot})\rangle,
\label{eq:PTanz}
\end{equation}
where the operator $\hat P^I_{MK}$ is the angular-momentum projector,
and $g^I_{K,\alpha}(\omega_{\rm rot})$ is the amplitude of superposition.
Here the cranked mean-field state, $|\Phi(\omega_{\rm rot})\rangle$,
with the rotational frequency (angular velocity), $\omega_{\rm rot}$,
is determined by the selfconsistent cranking procedure~\cite{RS80}.
There is no principle like the Galilean invariance for the rotational motion,
and the configuration-mixing with respect to the cranking frequency
in Eq.~(\ref{eq:PTanz}) should be evaluated numerically,
as it will be discussed in more detail below in Sec.~\ref{sec:frame}.
The same method has been also applied recently for the GCM calculation
with respect to the $(\beta,\gamma)$ collective coordinates
in Ref.~\cite{BRE15}.

In the present work, we apply the framework
to the superdeformed rotational band,
which is one of the most striking rotational motions in nuclei,
see e.g. Refs.~{\cite{SDtwin,SDrevA,SDrevB,SDrevC}.
With this application we would like to demonstrate
the importance of multicranked configuration-mixing
especially for high-spin states;
proper description of the moment of inertia for superdeformed states
cannot be obtained by the angular-momentum-projection calculation
from a single mean-field state.
The superdeformed band is best suited for this investigation
because long rotational sequences have been measured
without any quasiparticle-alignments in many cases.
Moreover, the influence of pairing correlations
is relatively small
on the moments of inertia of the superdeformed states;
for the normal deformed states studied in the previous work~\cite{STS16}
the reduction of moments of inertia due to the static pairing correlations
is very large, the reduction factor being 1/2 to 1/3 as is well-known,
and it is not so easy to identify the importance of
multicranked configuration-mixing.

The yrast superdeformed bands of two even-even nuclei,
$^{152}$Dy and $^{194}$Hg,
are selected as representative examples
in the $A\approx 150$ and $A\approx 190$ regions.
It has been known that the behaviors of
the dynamic moments of inertia~\cite{BM81}, $\Jtwom$,
are different for the superdeformed bands in the two mass regions.
The reason for selecting these two nuclei is that
the linking transitions were measured and the spin assignments were given.
Therefore we can study both $\Jonem$ and $\Jtwom$ moments of inertia
as functions of angular momentum.
Another purpose of the present paper is to look for
the main reason of the difference in these two mass regions;
the effect of pairing correlations is studied for this purpose
employing the cranked mean-field states obtained by
the variation after number projection approach.
Then the particle-number projection as well as the angular-momentum projection
have been carried out for the first time in this type of
multicranked configuration-mixing calculations.

The paper is organized as follows.
We briefly explain the basic formulation
of the actual procedure in Sec.~\ref{sec:frame}.
The results of calculation are presented in Sec.~\ref{sec:results},
where the importance of multicranked configuration-mixing is
discussed for typical examples of the yrast superdeformed bands
in the $^{152}$Dy and $^{194}$Hg nuclei.
A part of the results of moments of inertia for $^{152}$Dy
was already presented in Ref.~\cite{STS15};
the process of configuration-mixing leading to the results
is investigated in more detail
in comparison with another nucleus $^{194}$Hg in this section.
Sec.~\ref{sec:concls} is devoted to conclusion.

\section{Theoretical framework}
\label{sec:frame}

\subsection{Multicranked configuration-mixing}
\label{sec:multi}

Practically we discretize the continuous cranking frequency
$\omega_{\rm rot}$ in Eq.~(\ref{eq:PTanz}),
as \{$\omega^{(n)}_{\rm rot}$; $n=1,2,\cdots,n_{\rm max}$\},
\begin{equation}
 |\Psi^I_{M,\alpha}\rangle =
 \sum_{n=1}^{n_{\rm max}} \sum_{K} g^I_{Kn,\alpha}\,
 \hat P^I_{MK}|\Phi(\omega_{\rm rot}^{(n)})\rangle,
\label{eq:proj}
\end{equation}
and obtain the configuration-mixing amplitudes,
$g^I_{Kn,\alpha}\equiv g^I_{K,\alpha}(\omega_{\rm rot}^{(n)})$,
by solving the so-called Hill-Wheeler equation, see e.g. Ref.~\cite{RS80},
\begin{equation}
 \sum_{K'n'}{\cal H}^I_{Kn,K'n'}\ g^I_{K'n',\alpha} =
 E^I_\alpha\,
 \sum_{K'n'}{\cal N}^I_{Kn,K'n'}\ g^I_{K'n',\alpha},
\label{eq:HW}
\end{equation}
where the Hamiltonian and norm kernels are defined by
\begin{equation}
 \left\{ \begin{array}{c}
   {\cal H}^I_{Kn,K'n'} \\ {\cal N}^I_{Kn,K'n'} \end{array}
 \right\} = \langle \Phi(\omega_{\rm rot}^{(n)}) |
 \left\{ \begin{array}{c}
   H \\ 1 \end{array}
 \right\} \hat{P}_{KK'}^I | \Phi(\omega_{\rm rot}^{(n')})\rangle.
\label{eq:kernels}
\end{equation}
If the particle-number projection is performed
on top of the angular-momentum projection (see below), the wave function
in Eqs.~(\ref{eq:PTanz}), (\ref{eq:proj}) and (\ref{eq:kernels})
is replaced as
\begin{equation}
 \hat P^I_{MK}|\Phi(\omega_{\rm rot})\rangle
 \rightarrow
 \hat P^I_{MK} \,\hat P^{N_0} \hat P^{Z_0}|\Phi(\omega_{\rm rot})\rangle,
\label{eq:addnp}
\end{equation}
where $\hat P^{N_0}$ and $\hat P^{Z_0}$ are
the neutron- and proton-number projectors fixing
the neutron and proton numbers to the desired values $N_0$ and $Z_0$,
respectively.
If the particle-number projection is not performed,
the number conservation is treated approximately
by replacing $H \rightarrow H-\lambda_\nu (N-N_0)-\lambda_\pi (Z-Z_0)$.
As for the neutron and proton chemical potentials
$\lambda_\nu$ and $\lambda_\pi$
we use those of the first state $|\Phi(\omega_{\rm rot}^{(1)})\rangle$.

We have recently developed an efficient method for
the angular-momentum-projection and
the configuration-mixing calculation~\cite{TS12}.
This method is fully utilized also in the present work.
More details of our theoretical framework can be found
in Refs.~\cite{TS12,STS15}.
As for an effective interaction in the Hamiltonian $H$,
we employ the Gogny force~\cite{DeGo80}
with the so-called D1S parametrization~\cite{D1S}
as in our previous works~\cite{STS15,TS16,STS16};
therefore, there is no adjustable parameter in the Hamiltonian.
This interaction has been utilized in many applications
of the Hartree-Fock-Bogolyubov (HFB) calculation and
various theoretical methods beyond it, see e.g. Ref.~\cite{Egido16}.

\subsection{Determination of the mean-field states}
\label{sec:detmf}

The mean-field state, $|\Phi(\omega_{\rm rot})\rangle$
with $\omega_{\rm rot}=\omega^{(n)}_{\rm rot}$,
is determined by the cranked HFB procedure
with the routhian (cranked Hamiltonian), $H-\omega_{\rm rot}J_y$,
i.e., by the variation,
\begin{equation}
\delta\langle\Phi(\omega_{\rm rot})|
 H-\omega_{\rm rot}J_y |\Phi(\omega_{\rm rot})\rangle =0.
\label{eq:cHFB}
\end{equation}
The selfconsistent mean-fields of the superdeformed nuclei studied
in the present work are axially-deformed in a good approximation
even at highest frequencies.
We choose the $z$ axis as the (approximate) symmetry axis
and the $y$ axis as a cranking axis,
as it can be seen in Eq.~(\ref{eq:cHFB}).
Namely, we consider the one-dimensional cranking in the present work.
The full three-dimensional cranking~\cite{KO81},
or the tilted-axis cranking~\cite{Fra93,Fra00}, is necessary
when deformation of the mean-field strongly breaks the axial symmetry.
The infinitesimal cranking for such a case
was worked out in Ref.~\cite{TS16}.

It should be mentioned that with this cranked HFB procedure
the pairing correlation vanishes suddenly at some critical frequency.
However, for the finite system like a nucleus,
the pairing phase-transition takes place gradually and
the effect of pairing fluctuations plays non-negligible roles
near and after the critical frequency, see e.g. Ref.~\cite{pairRMP}.
In this reference, the pairing fluctuations calculated by
the random-phase approximation (RPA) method have been investigated
at the high-spin states, and shown to systematically improve
the agreement of the routhians and alignments with experimental data;
see also Refs.~\cite{SDB00,ADF00}.

The same methodology has been successfully applied to investigate
the $\Jonem$ and $\Jtwom$ moments of inertia of the superdeformed bands
in the $A\approx 150$ region~\cite{SVB90}, where the Nilsson-Strutinsky
method with the schematic monopole pairing interaction has been utilized.
It is, however, noted that the angular-momentum projection from
the RPA-correlated state is difficult to perform,
though not impossible; see e.g. Ref.~\cite{FR85}.
An alternative method to incorporate the pairing fluctuations
into the mean-field is the method of
variation after (particle-)number projection (VANP),
see e.g. Ref.~\cite{RS80}.
The effect of pairing fluctuations calculated
by the RPA and the VANP methods has been compared in Ref.~\cite{SB90}
and it has been confirmed that the two methods give
very similar results for observable quantities,
see also the discussion in Ref.~\cite{YRS90}.
In Ref.~\cite{NWJ89} the effect of pairing correlation has been
studied for the superdeformed bands in the $A\approx 150$ region
using the number projection method,
where the Woods-Saxon-Strutinsky calculation
with the schematic monopole pairing interaction has been employed,
and the pairing-gap parameter has been taken as a variational parameter.
Similar improvement over the simple mean-field approximation
has been obtained in these Refs.~\cite{SVB90,NWJ89}.

In relation to these developments,
it may be worthwhile mentioning that microscopic mean-field calculations
employing the Skyrme force with various density-dependent zero-range
pairing interactions have been performed for superdeformed nuclei
in the $A\approx 190$ region~\cite{THB95}
and in the $A\approx 150$ region~\cite{BFH96} with good agreements.
The same line of investigation but by using the Gogny force
has been reported, see e.g. Refs.~\cite{GDB94,VER00};
in the latter reference~\cite{VER00}
the effect of approximate number projection on the superdeformed states
has been also investigated.
The relativistic mean-field method has been also successfully applied
to the high-spin superdeformed rotational bands
in the $A\approx 140-150$ mass region, see e.g. Ref.~\cite{AKR96}.

To incorporate the effect of pairing fluctuations,
we also present the results of calculation, where the mean-field
state, $|\Phi(\omega_{\rm rot})\rangle$, is determined
by the VANP method, instead of the cranked HFB method in Eq.~(\ref{eq:cHFB}),
i.e., by the variation,
\begin{equation}
\delta\frac{\langle\Phi(\omega_{\rm rot})|
 (H-\omega_{\rm rot}J_y) \hat P^{N_0} \hat P^{Z_0}
 |\Phi(\omega_{\rm rot})\rangle}
 {\langle\Phi(\omega_{\rm rot})|\hat P^{N_0} \hat P^{Z_0}
 |\Phi(\omega_{\rm rot})\rangle}=0.
\label{eq:cVANP}
\end{equation}
In the present work the Gogny D1S effective interaction is used
and the variation with respect to the full-HFB amplitudes should be performed.
One of the common method is the gradient method, see e.g. Ref.~\cite{RS80},
but it takes a lot of iterations to achieve precise convergence.
An efficient method by utilizing diagonalization of
the number-projected quasiparticle Hamiltonian
has been developed in Ref.~\cite{SR00};
we make its full use to obtain the cranked VANP mean-field state
in Eq.~(\ref{eq:cVANP}).
With the VANP method, the particle-number projection should be performed
on top of the angular-momentum projection
for multicranked configuration-mixing, see Eq.~(\ref{eq:addnp}).

Since the deformation is axially-symmetric in a good approximation,
we define the $\lambda$-pole deformation parameter
of the calculated mean-field defined as usual by~\cite{IMY02}
\begin{equation}
\beta_{\lambda}\equiv
 \frac{4\pi}{3}\, \frac{\displaystyle
 \biggl\langle \sum_{i=1}^{A}(r^\lambda Y_{\lambda 0})^{}_i \biggr\rangle}
 {A\, \bar{R}^\lambda},\qquad\mbox{with}\quad \bar{R}=
 \sqrt{ \frac{5}{3A}\biggl\langle \sum_{i=1}^{A}r_i^2 \biggr\rangle},
\label{eq:defparm}
\end{equation}
and the average pairing gap by~\cite{BRRM00}
\begin{equation}
 \Delta \equiv
 \frac{\displaystyle -\sum_{a>b}\Delta_{ab}\kappa^*_{ab}}
 {\displaystyle \sum_{a>b}\kappa^*_{ab}},\qquad\mbox{with}\quad
  \Delta_{ab}= \sum_{c>d}\bar{v}_{ab,cd}\,\kappa_{cd},
\label{eq:gapparm}
\end{equation}
where the quantity $\kappa_{ab}$ is the abnormal density matrix
(the pairing tensor) and $\Delta_{ab}$ is the matrix element
of the pairing potential with the anti-symmetrized matrix element
$\bar{v}_{ab,cd}$ of the general two-body interaction,
see e.g. Ref.~\cite{RS80}.

\subsection{Two moments of inertia}

Although it is a textbook matter,
we here summarize the expressions of the two moments of inertia,
that is, the kinematic and dynamic ones~\cite{BM81}, for completeness.
For the spectrum of simple one-dimensional rotation $E(I)$,
the rotational frequency $\omega_{\rm rot}$ is defined by
\begin{equation}
 \omega_{\rm rot}=\frac{dE}{dI},
\label{eq:omega}
\end{equation}
which determine the $\omega_{\rm rot}-I$ relation $I(\omega_{\rm rot})$.
Then, the routhian, i.e.,
the energy in the rotating frame $E'(\omega_{\rm rot})$,
is given by the Legendre transformation,
\begin{equation}
 E'(\omega_{\rm rot})=E\bigl(I(\omega_{\rm rot})\bigr)
 -\omega_{\rm rot}I(\omega_{\rm rot}).
\label{eq:routh}
\end{equation}
With these definitions, the two moments of inertia,
$\Jtwom$ and $\Jonem$,
are expressed in various equivalent ways by
\begin{equation}
 \Jonem = I\left(\frac{dE}{dI}\right)^{-1}
 =\frac{I}{\omega_{\rm rot}}=
 -\frac{1}{\omega_{\rm rot}}\frac{dE'}{d\omega_{\rm rot}},
\label{eq:defJ1}
\end{equation}
\begin{equation}
 \Jtwom = \left(\frac{d^2E}{dI^2}\right)^{-1}
 =\frac{dI}{d\omega_{\rm rot}}=
 -\frac{d^2E'}{d\omega^2_{\rm rot}}.
\label{eq:defJ2}
\end{equation}

\section{Results of calculation}
\label{sec:results}

\subsection{Details of calculation}
\label{sec:detcal}

In the mean-field calculation such as the cranked HFB or VANP and
the subsequent angular-momentum-projection calculation,
the isotropic harmonic oscillator basis expansion is employed,
where all the basis states with the oscillator quantum numbers
$(n_x,n_y,n_z)$ satisfying $n_x+n_y+n_z \le N_{\rm osc}^{\rm max}=12$
are retained.
The value of canonical basis cut-off factor to define
the effective quasiparticle space
is taken to be $10^{-6}$ in the same way as in Ref.~\cite{TS12}.
The value of norm cut-off factor for solving the Hill-Wheeler equation,
see e.g. Ref.~\cite{RS80}, is chosen to be $10^{-12}\sim 10^{-9}$
in order to obtain as continuous rotational bands as possible~\cite{STS15,STS16}.

The maximum values of the angular-momentum and its projection to
the (approximate) symmetry axis are taken to be
$I_{\rm max}=62$ and $K_{\rm max}=22$ for $^{152}$Dy,
and $I_{\rm max}=52$ and $K_{\rm max}=22$ for $^{194}$Hg.
Note that the cranking procedure with high rotational frequency
causes considerable $K$-mixing
although the deformation is approximately axially-symmetric, and,
therefore, $K_{\rm max}$ should not be very small.
We have confirmed that the selected values above are enough
for the present calculation.
As for the numbers of integration-mesh points for the Euler angles
($\alpha,\beta,\gamma$) in the calculation of angular-momentum projector,
rather large values especially for the $\beta$ integration are necessary
to obtain precise energy spectrum up to high-spin states like $I\approx 60$.
We have used $N_\beta=130$ and $N_\alpha=N_\gamma=60$
after confirming accuracy of the results.
To perform the VANP calculation in Eq.~(\ref{eq:cVANP})
the particle-number projector should be applied,
for which the number of integration-mesh points
for the gauge angle $\phi$ has been taken to be $N_\phi=7$.

As for the number of mean-field states of multicranked configuration-mixing
in Eq.~(\ref{eq:proj}), we use $n_{\rm max}=4$ in the present work.
A larger value of $n_{\rm max}$ is preferable, but the numerical efforts
to perform the angular-momentum projection are too much
with relatively large numbers of integration-mesh points
for the Euler angles being employed in the present calculation;
note that the numerical cost increases in proportion to $n^2_{\rm max}$.  
With the mean-field states obtained by the VANP method,
we carry out the particle-number projection
on top of the angular-momentum projection
as is explained in Sec.~\ref{sec:detmf};
then the numerical cost is $N_\phi\,(=7)$ times larger.

The discretized points of the rotational frequency,
$\omega^{(n)}_{\rm rot}$ ($n=1,\cdots,n_{\rm max}=4$),
can be chosen rather arbitrarily; in Ref.~\cite{STS15} it is discussed
that the final results of configuration-mixing do not depend on this choice
when enough number of points are employed.
We select the first and last points,
$\omega^{(1)}_{\rm rot}$ and $\omega^{(4)}_{\rm rot}$,
and other points are determined to form the equidistant-mesh.
Some trial-and-error has been done to obtain the smooth rotational band,
which is necessary to calculate kinematic and dynamic moments of inertia.
In the following, we mainly discuss the $\Jtwom$ moment of inertia
evaluated with the discrete ${\mit\Delta}I=\pm 2$ rotational spectrum, $E(I)$,
\begin{equation}
 \Jtwom(I) = \frac{4\hbar^2} {E(I+2)+E(I-2)-2E(I)}
 = \frac{4\hbar^2} {E_\gamma(I+1)-E_\gamma(I-1)},
\label{eq:J2}
\end{equation}
for the experimental data and for the calculated results of projection,
where $E_\gamma(I)\equiv E(I+1)-E(I-1)$ is the $\gamma$-ray energy of
the $I+1 \rightarrow I-1$ transition.
Note that these quantities can be evaluated only with
the $\gamma$-ray energies without the spin-assignment,
which is often missing for superdeformed rotational bands.
We also discuss the $\Jonem$ moment of inertia,
\begin{equation}
 {\cal J}^{(1)}(I)
 = \frac{(2I+1)\hbar^2}{E(I+1)-E(I-1)}
 = \frac{(2I+1)\hbar^2}{E_\gamma(I)},
\label{eq:J1}
\end{equation}
which requires the spin-assignment to calculate.
The $\Jtwom$ moment of inertia calculated
within the cranked HFB approximation,
\begin{equation}
 {\cal J}^{(2)}(\omega_{\rm rot})
 =\frac{d}{d \omega_{\rm rot}}
 \langle \Phi(\omega_{\rm rot}) |J_y| \Phi(\omega_{\rm rot}) \rangle,
\label{eq:J2mf}
\end{equation}
as a function of semiclassical spin value defined by
\begin{equation}
 I(\omega_{\rm rot}) \equiv
 \langle \Phi(\omega_{\rm rot}) |J_y| \Phi(\omega_{\rm rot}) \rangle
 -\half\hbar,
\label{eq:Imf}
\end{equation}
or within the cranked VANP approximation,
\begin{equation}
 {\cal J}^{(2)}(\omega_{\rm rot})
 =\frac{d}{d \omega_{\rm rot}}
 \frac{\langle \Phi(\omega_{\rm rot}) |
 J_y \,\hat P^{N_0} \hat P^{Z_0}| \Phi(\omega_{\rm rot}) \rangle}
 {\langle\Phi(\omega_{\rm rot})|\hat P^{N_0} \hat P^{Z_0}
 |\Phi(\omega_{\rm rot})\rangle},
\label{eq:J2mfnp}
\end{equation}
as a function of
\begin{equation}
 I(\omega_{\rm rot}) \equiv
 \frac{\langle \Phi(\omega_{\rm rot}) |
 J_y \,\hat P^{N_0} \hat P^{Z_0}| \Phi(\omega_{\rm rot}) \rangle}
 {\langle\Phi(\omega_{\rm rot})|\hat P^{N_0} \hat P^{Z_0}
 |\Phi(\omega_{\rm rot})\rangle} -\half\hbar,
\label{eq:Imfnp}
\end{equation}
are also compared with the result of projected configuration-mixing
in the following discussion.

The experimental data are taken from Ref.~\cite{SDdata}.
As for the reference of deformation for superdeformed bands,
see e.g. Refs.~\cite{WD92,WD95}.
Note, however, that the deformation parameter studied
in these references is that of the mean-field potential
with Woods-Saxon shape.  Their values are systematically smaller than
those of the deformation parameter determined according to
the density distribution in Eq.~(\ref{eq:defparm}),
see e.g. Ref.~\cite{DNO84}.

\subsection{Superdeformed band in $^{\bf 152}$Dy}
\label{sec:Dy152}

We first investigate the yrast superdeformed band of the $^{152}$Dy nucleus,
which has been identified as a first high-spin superdeformed
band~\cite{SDtwin} in the $A\approx 150$ region
and the spin-assignment has been given afterward~\cite{Dy152SD}.
In the present work,
we generate the cranked mean-field states
employed for the projected configuration-mixing
by choosing four rotational frequencies,
$\hbar\omega_{\rm rot}=0.30,\,0.45,\,0.60,\,0.75$ MeV,
which roughly cover the range of the observed rotational band in $^{152}$Dy.

\subsubsection{Projection from the mean-field determined
by cranked HFB method}
\label{sec:Dy152HFB}

Our Gogny HFB calculation gives a superdeformed minimum with deformation
$\beta_2=0.715$ at zero rotational frequency
with very weak pairing correlations in $^{152}$Dy.
The pairing correlations quickly vanish and the mean-field states
are non-superconducting at $\hbar\omega_{\rm rot} \ge 0.3$ MeV,
with which the multicranked configuration-mixing is performed
after projection.  The deformation is almost constant keeping
the axial-symmetry very well up to high rotational frequency;
the calculated values of deformation parameter
are $\beta_2=0.713$ and $0.696$ at
$\hbar\omega_{\rm rot}=0.30$ and $0.75$ MeV, respectively.
Note that this deformation reproduces
the observed $B(E2)$ values in $^{152}$Dy very well
as it was confirmed in our previous work~\cite{STS15}.
In Fig.~\ref{fig:dymoi} the calculated $\Jtwom$ moment of inertia
is compared with the experimental one.
Both the calculated and experimental $\Jtwom$ moments of inertia
are almost constant or only gradually decrease
as functions of angular momentum,
and the calculated one slightly overestimates the experimental one.
The result of configuration-mixing is very similar to
that of Ref.~\cite{STS15}, where the four cranked mean-field states
with different frequency points,
$\hbar\omega_{\rm rot}=0.01,\,0.24,\,0.47,\,0.70$ MeV,
have been utilized instead.
This shows, again, that the result is almost independent of
the choice of actual mesh points of
the cranking frequency in Eq.~(\ref{eq:PTanz}), or Eq.~(\ref{eq:proj}),
with relatively small number of them ($n_{\rm max}=4$ in the present case).
The small norm cut-off factor $10^{-12}$ can be used in this calculation.
\begin{figure}[!htb]
\begin{center}
\includegraphics[width=75mm]{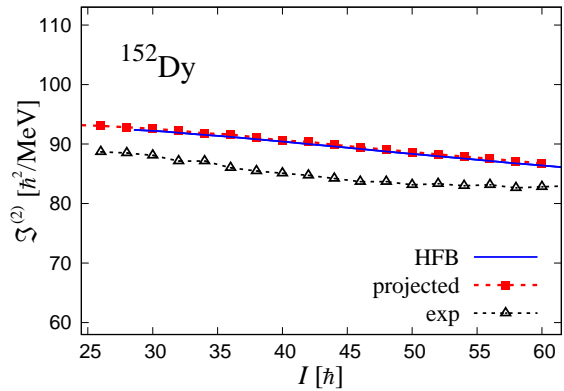}
\end{center}
\vspace*{-8mm}
\caption{(Color online)
Calculated $\Jtwom$ moment of inertia
as a function of angular momentum for $^{152}$Dy,
where the result of projected multicranked configuration-mixing
is drawn with symbols and that of cranked HFB calculated
with Eq.~(\ref{eq:J2mf}) by the solid line.
The experimental one is also included.
}
\label{fig:dymoi}
\end{figure}

The results of multicranked configuration-mixing after projection
and of the simple cranked HFB mean-field approximation
shown in Fig.~\ref{fig:dymoi} are very similar.
One may think that it is natural, but this is totally non-trivial
and a consequence of the multicranked configuration-mixing
as is shown in Fig.~\ref{fig:dymoimix},
where four $\Jtwom$ moments of inertia calculated from
the spectra projected from a single mean-field state
at each rotational frequency are depicted
in addition to the final result of configuration-mixing.
It can be seen that the values of these $\Jtwom$ moments of inertia are
very similar and about $20-25$\% smaller than
the one obtained by the result of configuration-mixing,
and, furthermore,
they decrease more rapidly as functions of angular momentum.
It seems that this is a general trend for the projected spectrum
from a single HFB mean-field state~\cite{STS15,STS16},
and indicates the importance of multicranked configuration-mixing
for the proper description of the moment of inertia
of the rotational band by the angular-momentum-projection method,
especially for high-spin states.
\begin{figure}[!htb]
\begin{center}
\includegraphics[width=75mm]{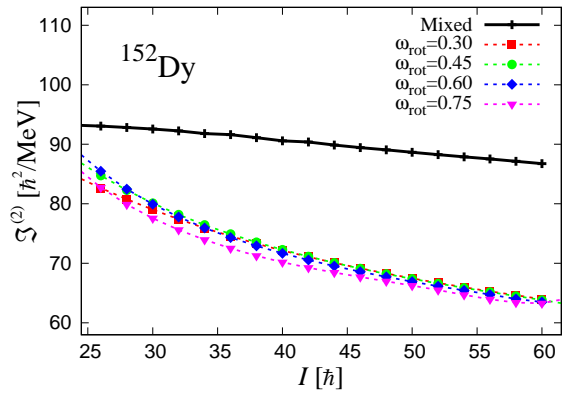}
\end{center}
\vspace*{-8mm}
\caption{(Color online)
$\Jtwom$ moments of inertia of projected spectra from
a single cranked HFB state at each rotational frequency for $^{152}$Dy.
The result of configuration-mixing is also included.
}
\label{fig:dymoimix}
\end{figure}

It may be worthwhile mentioning that the $\Jtwom$ moment of inertia
calculated by the projection from a single mean-field state
shown in Fig.~\ref{fig:dymoimix} corresponds to
the so-called Yoccoz inertia,
while the one calculated within the cranked HFB mean-field approximation,
cf. Eq.~(\ref{eq:J2mf}),
which agrees with the final result of projected configuration-mixing as shown
in Fig.~\ref{fig:dymoi}, corresponds to
the Thouless-Valatin inertia, see Ref.~\cite{RS80}.
Thus, this result shows that the Yoccoz inertia is considerably smaller
than the Thouless-Valatin inertia at least for
the superdeformed rotational band at high-spin states;
it is true not only for $^{152}$Dy but also for $^{194}$Hg
as will be shown in Figs.~\ref{fig:hgmoi} and~\ref{fig:hgmoimix} below.

In order to see why the result of configuration-mixing gives
considerably larger value for $\Jtwom$ moment of inertia,
the calculated spectra obtained by the projection
from a single cranked HFB state at each rotational frequency
as well as the result of the configuration-mixing
are shown in Fig.~\ref{fig:dyenergy},
where the reference energy, $I(I+1)/190$ MeV, is subtracted.
The $\Jtwom$ moment of inertia is the reciprocal of curvature
of the spectral curve as a function of angular momentum,
cf. Eq.~(\ref{eq:defJ2}).
The result of configuration-mixing looks naturally like
the envelope curve of a family of four spectral curves
corresponding to those obtained with different frequencies,
and consequently the curvature reduces from those of a family of curves.
Note that each spectral curve obtained by projection
from a single cranked HFB state with
$\omega_{\rm rot}=\omega_{\rm rot}^{(n)}$ $(n=1,\cdots,4)$
comes in contact with this envelope-like curve
at the spin value close to the cranked angular-momentum
 $I\approx\langle\Phi(\omega_{\rm rot}^{(n)}) |
  J_y| \Phi(\omega_{\rm rot}^{(n)}) \rangle$.
The resultant spectrum of configuration-mixing is very similar
to the one calculated by the cranked HFB
as in the case of $\Jtwom$ moments of inertia shown in Fig.~\ref{fig:dymoi}.
In this way, the considerable increase of the $\Jtwom$ moment of inertia
caused by the multicranked configuration-mixing can be naturally understood.
\begin{figure}[!htb]
\begin{center}
\includegraphics[width=75mm]{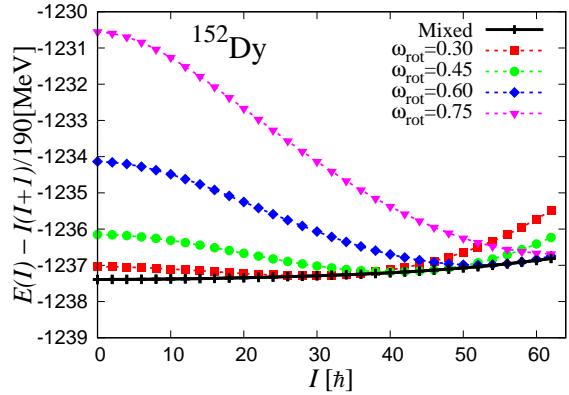}
\end{center}
\vspace*{-8mm}
\caption{(Color online)
Energy spectra of simple projections from
a single cranked HFB state at each rotational frequency and
that of the resultant configuration-mixing for $^{152}$Dy.
The reference energy, $I(I+1)/190$ MeV, is subtracted.
}
\label{fig:dyenergy}
\end{figure}
\begin{figure}[!htb]
\begin{center}
\includegraphics[width=75mm]{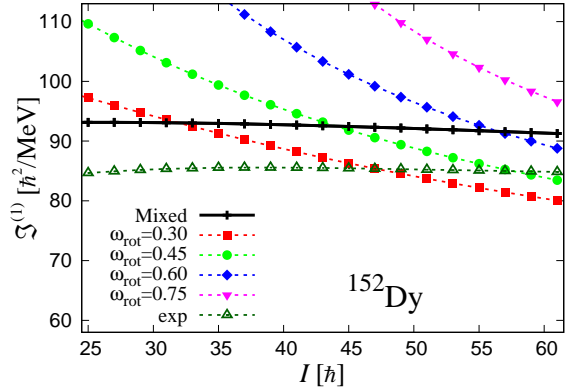}
\end{center}
\vspace*{-8mm}
\caption{(Color online)
$\Jonem$ moments of inertia corresponding to
the spectra in Fig.~\ref{fig:dyenergy} for $^{152}$Dy.
The experimental one is also included.
}
\label{fig:dyj1}
\end{figure}

In addition to the $\Jtwom$ moment of inertia in Fig.~\ref{fig:dymoi},
the $\Jonem$ moment of inertia is also useful for studying
the properties of high-spin rotational bands.
The calculated $\Jonem$ moments of inertia
corresponding to Fig.~\ref{fig:dyenergy} are shown in Fig.~\ref{fig:dyj1},
where the experimental one is also included.
As seen from the figure, the value of the $\Jonem$ moment of inertia
is larger for the spectrum obtained by the projection from the mean-field
state with higher rotational frequency.
Those calculated by the projection from a single mean-field state
are larger and decrease more rapidly than
the corresponding $\Jtwom$ moments of inertia in Fig.~\ref{fig:dymoimix}.
However, the value of the result of configuration-mixing
is almost constant in agreement with the trend of the experimental data,
although the calculated value of $\Jonem$
is considerably (about 10\%) overestimated.
In this way, the projected spectrum from a single mean-field state
does not give good description of the high-spin rotational band,
and the multicranked configuration-mixing is crucial
to obtain the correct magnitudes of both dynamic and kinematic
moments of inertia for the superdeformed band in $^{152}$Dy.

\subsubsection{Projection from the mean-field determined
by cranked VANP method}
\label{sec:Dy152VANP}

\begin{figure}[!htb]
\begin{center}
\includegraphics[width=75mm]{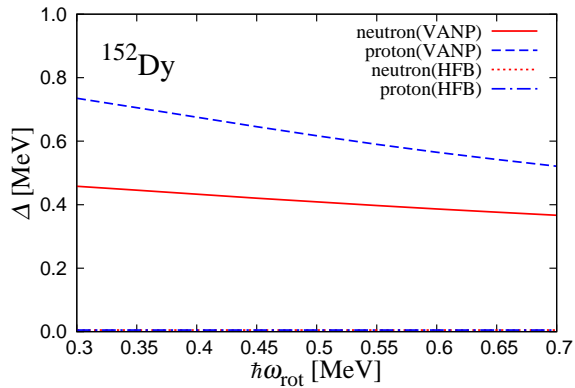}
\end{center}
\vspace*{-8mm}
\caption{(Color online)
Average pairing gaps as functions of the rotational frequency
for $^{152}$Dy, defined by Eq.~(\ref{eq:gapparm}) and
calculated by the cranked HFB and VANP methods.
Both neutron and proton gaps obtained by the HFB method
vanish in the frequency range shown.
}
\label{fig:dygap}
\end{figure}
As it is discussed in Sec.~\ref{sec:detmf},
the pairing phase-transition occurs suddenly at low spins
with the static mean-field approximation like the HFB method.
However, the effect of pairing fluctuations remains
at rather high-spin states~\cite{pairRMP,YRS90}.
An efficient method to take it into account is the VANP method,
which determines the mean-field state according to Eq.~(\ref{eq:cVANP}).
With the cranked VANP method, the obtained mean-field states
have almost the same deformation as in the case of the cranked HFB method,
$\beta_2=0.713$ and $0.698$ at
$\hbar\omega_{\rm rot}=0.30$ and $0.75$ MeV, respectively.
They are, however, in the superconducting phase
with relatively weak pairing correlations,
as is shown in Fig.~\ref{fig:dygap},
where the calculated average pairing gaps
by Eq.~(\ref{eq:gapparm}) are depicted.
The average pairing gaps for the neutron and proton
obtained by the VANP method gradually decrease
as functions of the rotational frequency
and never vanish within the frequency range under consideration.
These results are consistent with those in Ref.~\cite{VER00}.

\begin{figure}[!htb]
\begin{center}
\includegraphics[width=75mm]{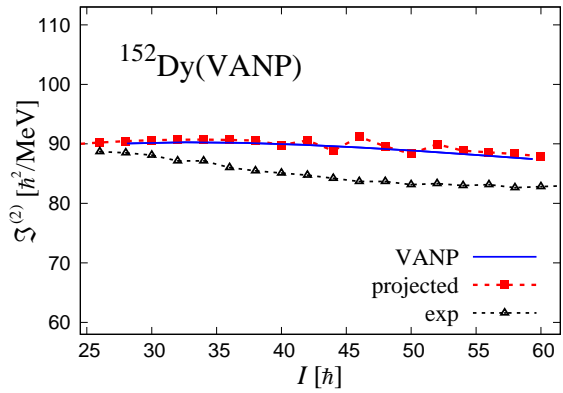}
\end{center}
\vspace*{-8mm}
\caption{(Color online)
Similar to Fig.~\ref{fig:dymoi} but the mean-field states
obtained by the VANP method are utilized
and the particle-number projection is also performed, cf. Eq.~(\ref{eq:addnp}).
Here the solid line is calculated by Eq.~(\ref{eq:J2mfnp}).
}
\label{fig:dynpmoi}
\end{figure}
Utilizing four mean-field states obtained by this VANP method
at the same cranking frequencies as in the case of the HFB method,
$\hbar\omega_{\rm rot}=0.30,\,0.45,\,0.60,\,0.75$ MeV,
the projected multicranked configuration-mixing has been carried out,
where the particle-number is also projected out to the desired number
for both the neutron and proton, see Eq.~(\ref{eq:addnp}).
A larger norm cut-off factor $10^{-9}$ is necessary to obtain
a smooth rotational band.
The resultant $\Jtwom$ moment of inertia is depicted
in Fig.~\ref{fig:dynpmoi}.
Apparently, the result does not change very much from
that with the cranked HFB method shown in Fig.~\ref{fig:dymoi},
although it is almost constant as a function of angular momentum
in contrast to the experimental data and
the discrepancy slightly increases at higher frequency.
The result of the simple cranked VANP approximation (the solid line)
in Eq.~(\ref{eq:J2mfnp}) is again very similar to
that of projected multicranked configuration-mixing
as in the case of the HFB mean-field states being employed.

The calculated $\Jtwom$ moment of inertia
by the projected configuration-mixing in Fig.~\ref{fig:dynpmoi}
shows some small irregularities.
This and other irregularities in the calculations seen in the present work
are due to the fact that a small norm state in the Hill-Wheeler equation
unfortunately comes in and/or goes out in the calculated spin range
even though a relatively small value of norm cut-off parameter
has been employed, in this case $10^{-9}$;
usually its effect is small but it can be visible
especially for the $\Jtwom$ moment of inertia
that is the quantity of second derivative,
see Eq.~(\ref{eq:defJ2}).

\begin{figure}[!htb]
\begin{center}
\includegraphics[width=75mm]{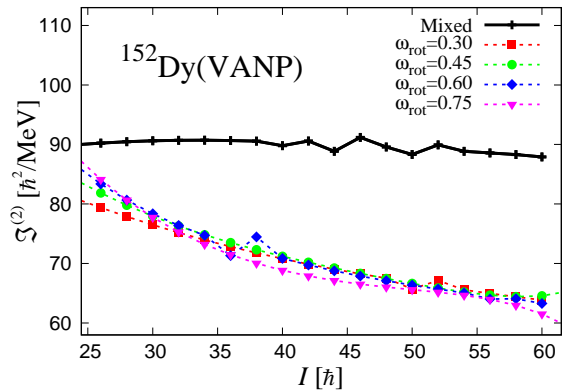}
\end{center}
\vspace*{-8mm}
\caption{(Color online)
Similar to Fig.~\ref{fig:dymoimix} but the mean-field states
obtained by the VANP method are utilized
and the particle-number projection is also performed, cf. Eq.~(\ref{eq:addnp}).
}
\label{fig:dynpmoimix}
\end{figure}
We show four $\Jtwom$ moments of inertia calculated from spectra
projected from a single mean-field state at each rotational frequency
in Fig.~\ref{fig:dynpmoimix} with the result of configuration-mixing:
They are rather similar to those obtained from the HFB mean-field states
shown in Fig.~\ref{fig:dymoimix},
even though the average pairing gaps remain finite
in the VANP mean-field states
contrary to the vanishing pairing gaps in the HFB states.
Thus, the effect of pairing fluctuations is not large for this nucleus.
The calculated spectra of projection from a single cranked VANP state
at each rotational frequency are displayed in Fig.~\ref{fig:dynpenergy}
with the resultant spectrum of configuration-mixing.
Note that the absolute energy of the projected configuration-mixing spectrum
using the VANP states is about 3.7 MeV smaller at $I\approx 0$
because the particle-number projection
is performed for both neutron and proton.
The $\Jonem$ moments of inertia corresponding to the spectra
in Fig.~\ref{fig:dynpenergy} are also displayed in Fig.~\ref{fig:dynpj1},
where the experimental one is also included.
The main features in Figs.~\ref{fig:dynpmoimix}$\sim$\ref{fig:dynpj1}
are not very different from the case utilizing the HFB mean-fields
in Figs.~\ref{fig:dymoimix}$\sim$\ref{fig:dyj1}.
\begin{figure}[!htb]
\begin{center}
\includegraphics[width=75mm]{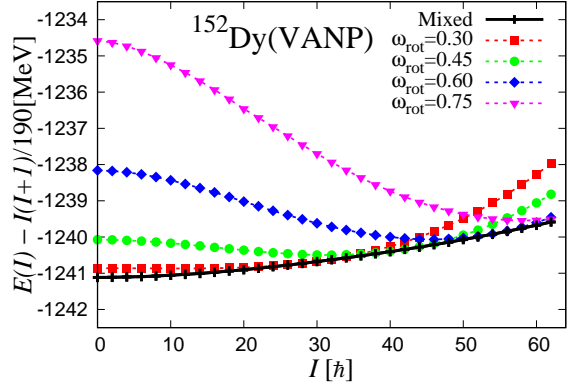}
\end{center}
\vspace*{-8mm}
\caption{(Color online)
Similar to Fig.~\ref{fig:dyenergy} but the mean-field states
obtained by the VANP method are utilized
and the particle-number projection is also performed, cf. Eq.~(\ref{eq:addnp}).
}
\label{fig:dynpenergy}
\end{figure}
\begin{figure}[!htb]
\begin{center}
\includegraphics[width=75mm]{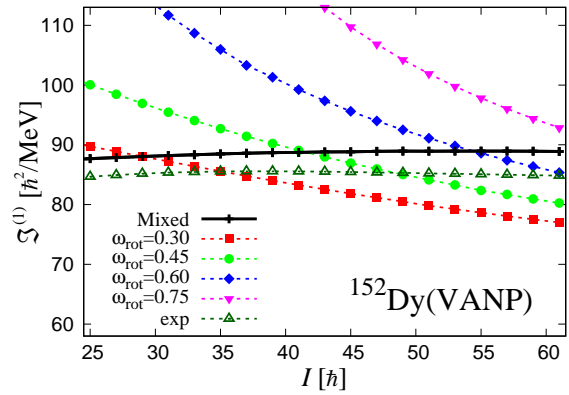}
\end{center}
\vspace*{-8mm}
\caption{(Color online)
Similar to Fig.~\ref{fig:dyj1} but the mean-field states
obtained by the VANP method are utilized
and the particle-number projection is also performed, cf. Eq.~(\ref{eq:addnp}).
}
\label{fig:dynpj1}
\end{figure}

The difference between the results using
the HFB and VANP mean-field states is that
the values of cranked angular-momentum,
see Eqs.~(\ref{eq:Imf}) and~(\ref{eq:Imfnp}),
are systematically smaller in the results of the VANP states.
In fact the resultant $\Jonem$ moment of inertia by the configuration-mixing
with the VANP method is  reduced from that with the HFB method,
and the agreement between the experimental data is  improved,
as it can be seen by comparing the $\Jonem$ moments of inertia
in Figs.~\ref{fig:dyj1} and~\ref{fig:dynpj1}.
The importance of this ``dealignment'' effect and the reduction
of $\Jonem$ moment of inertia caused by the pairing fluctuations
has been systematically investigated for
normal deformed nuclei in Ref.~\cite{pairRMP}
and for superdeformed nuclei in Refs.~\cite{SVB90,NWJ89}.
It can be easily understood by the fact
that the correlation routhian induced by the pairing fluctuations,
which is always negative,
is increasing as a function of the rotational frequency
and vanishes at infinite frequency.
Then the correction due to the pairing fluctuations
for $\Jonem$ is always negative, cf. Eq.~(\ref{eq:defJ1}),
in agreement with the analysis of Refs.~\cite{SVB90,NWJ89}.
For the $\Jtwom$ moment of inertia, the sign of correction term
changes at the inflection point of the correlation routhian,
because the $\Jtwom$ moment of inertia is defined by the second derivative,
see e.g. Fig.~9 of Ref.~\cite{SVB90}.
Comparing the results of configuration-mixed spectra employing
the HFB and VANP mean-field states,
the inflection point of the correlation routhian exists
at a rather high rotational frequency
like $\hbar\omega_{\rm rot}\approx 0.5$ MeV,
and the correction to $\Jtwom$ is negative
before this frequency and positive after it,
although the magnitude of correction to $\Jtwom$ is rather small;
this result is slightly different from that in Ref.~\cite{SVB90}.

\subsection{Superdeformed band in $^{\bf 194}$Hg}
\label{sec:Hg194}

As an example of superdeformed bands in the $A\approx 190$ mass region,
we take the yrast superdeformed band of the $^{194}$Hg nucleus,
which has been observed in an early stage of superdeformation-hunting
in this region~\cite{Hg194sp},
and the spin-assignment has been given afterward~\cite{Hg194SD}.
We have performed the multicranked configuration-mixing calculation
using four cranked mean-field states at the rotational frequencies,
$\hbar\omega_{\rm rot}=0.10,\,0.23,\,0.36,\,0.49$ MeV,
which cover the spin-range of the measured rotational band in $^{194}$Hg.

\begin{figure}[!htb]
\begin{center}
\includegraphics[width=75mm]{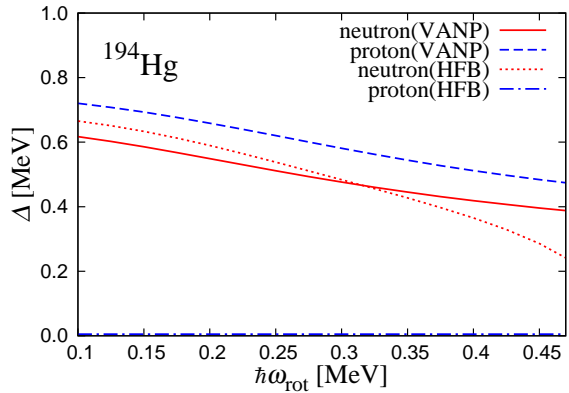}
\end{center}
\vspace*{-8mm}
\caption{(Color online)
Average pairing gaps as functions of the rotational frequency
for $^{194}$Hg,
defined by Eq.~(\ref{eq:gapparm}) and
calculated by the cranked HFB and VANP methods.
Proton gap obtained by the HFB method vanishes
in the frequency range shown.
}
\label{fig:hggap}
\end{figure}
Generally speaking, the pairing correlations are weak
due to the relatively large shell gaps
resulting from the (approximate) rational axis ratio 2:1
for the long- to short-axis in the superdeformed states~\cite{BM75}.
The values of deformation for the superdeformed states
in the $A\approx 190$ region are smaller than
those in the $A\approx 150$ region, see e.g. Ref.~\cite{Chas89},
and the shell gaps in the $A\approx 190$ region
are suggested to be slightly smaller.
Because of this, the pairing correlations are expected to be
relatively stronger for nuclei in the $A\approx 190$ region than
those in the $A\approx 150$ region, see e.g. Ref.~\cite{BFH90}.
The calculated average pairing gaps for $^{194}$Hg
are shown in Fig.~\ref{fig:hggap}
as functions of the rotational frequency.
The proton gap obtained by the HFB method vanishes already
at $\hbar\omega_{\rm rot}=0.10$ MeV,
while the neutron gap remains at higher frequency.
As in the case of $^{152}$Dy shown in Fig.~\ref{fig:dygap},
both pairing gaps for both neutron and proton
obtained by the VANP method are finite and gradually decrease
as functions of the rotational frequency.
These results are very similar to those in Ref.~\cite{VER00}.
The neutron gap with the HFB method decreases more rapidly
than that with the VANP method.
The values of the average gaps in $^{194}$Hg and $^{152}$Dy
calculated with the VANP method are similar for protons,
and the value in $^{194}$Hg is larger for neutrons.
Considering the general mass dependence
of the pairing gap, $\propto 1/\sqrt{A}$,
the pairing correlations deduced from the calculated pairing gaps
are stronger in $^{194}$Hg than in $^{152}$Dy especially for neutrons.

\subsubsection{Projection from the mean-field determined
by cranked HFB method}
\label{sec:Hg194HFB}

Our Gogny HFB calculation gives a superdeformed minimum
in $^{194}$Hg with $\beta_2=0.548$ and $0.529$ at
$\hbar\omega_{\rm rot}=0.10$ and $0.49$ MeV, respectively.
Again, the axial-symmetry is kept very well up to high rotational frequency.
The calculated $\Jtwom$ moment of inertia by the configuration-mixing
using the four cranked HFB mean-fields states
is shown in Fig.~\ref{fig:hgmoi},
where the result of simple HFB approximation, cf. Eq.~(\ref{eq:J2mf}),
and the experimental one are also included.
The small norm cut-off factor $10^{-12}$ works in this calculation.
In contrast to the case of $^{152}$Dy, where the $\Jtwom$ moment of
inertia is almost constant or even gradually decreases,
the calculated $\Jtwom$ moment of inertia for $^{194}$Hg increases
as a function of angular momentum,
which clearly shows the importance of the pairing correlation.
However, the amount of increase is considerably smaller
in comparison with the experimental data.
It should be mentioned that the result of
the HFB mean-field approximation (the solid line)
is very similar to that calculated by
the multicranked configuration-mixing
just like the case of $^{152}$Dy.
\begin{figure}[!htb]
\begin{center}
\includegraphics[width=75mm]{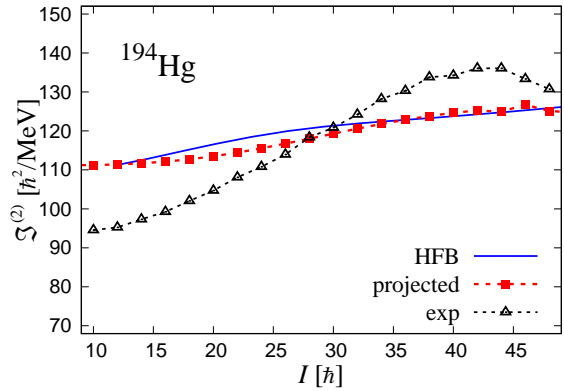}
\end{center}
\vspace*{-8mm}
\caption{(Color online)
Calculated $\Jtwom$ moment of inertia
as a function of angular momentum for $^{194}$Hg,
where the result of projected multicranked configuration-mixing
is drawn with symbols and that of cranked HFB mean-field
with Eq.~(\ref{eq:J2mf}) by the solid line.
The experimental one is also included.
}
\label{fig:hgmoi}
\end{figure}

In order to see the effect of configuration-mixing,
four $\Jtwom$ moments of inertia calculated from the spectra
obtained by the projection from a single cranked HFB state
at each rotational frequency are displayed in Fig.~\ref{fig:hgmoimix},
where the final result of configuration-mixing is also included.
As in the case of $^{152}$Dy in Fig.~\ref{fig:dymoimix},
they take similar values except for at lower spin values $I \ltsim 20$,
and they gradually decrease as functions of angular momentum.
The average values of four $\Jtwom$ moments of inertia
at high-spin states calculated by the projection
from a single cranked mean-fields state are considerably smaller
than the value obtained by the final configuration-mixing.
Moreover, the dependence on the angular momentum completely changes
as a result of configuration-mixing for $^{194}$Hg,
which leads to the increase in accordance with the experimental data.
\begin{figure}[!htb]
\begin{center}
\includegraphics[width=75mm]{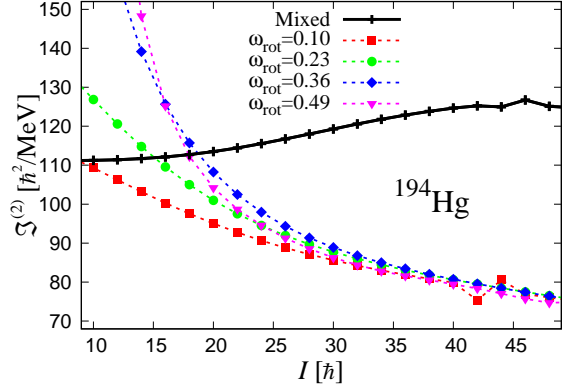}
\end{center}
\vspace*{-8mm}
\caption{(Color online)
$\Jtwom$ moments of inertia of projected spectra from
a single cranked HFB state at each rotational frequency for $^{194}$Hg.
The result of configuration-mixing is also included.
}
\label{fig:hgmoimix}
\end{figure}

Furthermore, the calculated spectra from a single cranked HFB state
at each rotational frequency are depicted in Fig.~\ref{fig:hgenergy}
with the result of final configuration-mixing;
the reference energy, $I(I+1)/240$ MeV, is subtracted for $^{194}$Hg.
The resultant spectral curve after the configuration-mixing
follows the envelope curve of the four spectral curves obtained
by the projection from a single mean-field state
with $\omega_{\rm rot}=\omega_{\rm rot}^{(n)}$ $(n=1,\cdots,4)$,
and comes in contact with each spectral curve at
 $I\approx\langle\Phi(\omega_{\rm rot}^{(n)}) |
  J_y| \Phi(\omega_{\rm rot}^{(n)}) \rangle$.
Consequently, its curvature is becoming smaller,
or the $\Jtwom$ moment of inertia is becoming larger,
as a result of configuration-mixing.
\begin{figure}[!htb]
\begin{center}
\includegraphics[width=75mm]{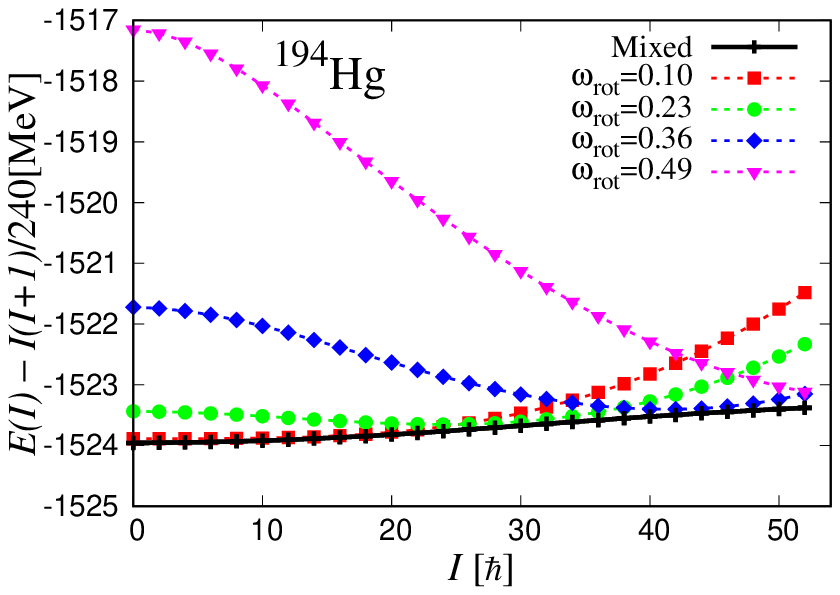}
\end{center}
\vspace*{-8mm}
\caption{(Color online)
Energy spectra of simple projections from
a single cranked HFB state at each rotational frequency and
that of the resultant configuration-mixing for $^{194}$Hg.
The reference energy, $I(I+1)/240$ MeV, is subtracted.
}
\label{fig:hgenergy}
\end{figure}
\begin{figure}[!htb]
\begin{center}
\includegraphics[width=75mm]{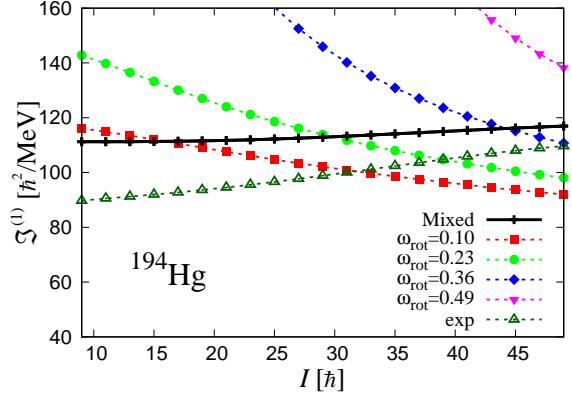}
\end{center}
\vspace*{-8mm}
\caption{(Color online)
$\Jonem$ moments of inertia corresponding to
the spectra in Fig.~\ref{fig:hgenergy} for $^{194}$Hg.
The experimental one is also included.
}
\label{fig:hgj1}
\end{figure}
In Fig.~\ref{fig:hgj1} the $\Jonem$ moments of inertia calculated
by the projection from a single mean-field state are compared with
those by the final configuration-mixing and by the experimental data.
As in the case of $^{152}$Dy, the four calculated $\Jonem$ moments of inertia
obtained by the projection from a single mean-field state quickly decrease
as spin increases.
The spin-dependence of the result of final configuration-mixing changes,
i.e., the resultant $\Jonem$ moment of inertia increases as spin,
which well corresponds well to the trend of the experimental data,
although the absolute value is considerably overestimated
compared to the experimental data.
Thus, the $\Jonem$ and $\Jtwom$ moments of inertia calculated
by the projection from a single mean-field state
with each rotational frequency are very different
from the experimentally measured moments of inertia
for both the absolute value and
the spin-dependence.
Again, the effect of multicranked configuration-mixing
is essential to understand the observed behavior
of the rotational spectrum and two moments of inertia.

\subsubsection{Projection from the mean-field determined
by cranked VANP method}
\label{sec:Hg194VANP}

In order to see the effect of dynamic pairing correlations,
we have performed the angular-momentum-projection calculations
employing the mean-field states obtained by the VANP method for $^{194}$Hg;
the particle-number projection is also performed in this case,
cf. Eq.~(\ref{eq:addnp}).
The calculated values of deformation parameter for superdeformed minimum
are $\beta_2=0.545$ and $0.530$ at
$\hbar\omega_{\rm rot}=0.10$ and $0.49$ MeV, respectively,
which are essentially the same as those calculated with the HFB method.
With the four mean-field states obtained by the cranked VANP method
at the same cranking frequencies as those by the cranked HFB method,
$\hbar\omega_{\rm rot}=0.10,\,0.23,\,0.36,\,0.49$ MeV,
the multicranked configuration-mixing calculation
has been carried out with the result shown in Fig.~\ref{fig:hgnpmoi}.
A larger norm cut-off factor $10^{-10}$ is necessary to obtain
smooth rotational band.
As it is clearly seen, the result employing the VANP mean-field states
is slightly changed from the one
employing the HFB mean-field states shown in Fig.~\ref{fig:hgmoi}:
The calculated $\Jtwom$ moment of inertia is reduced at lower spin
and is increased at higher spin
in comparison with the one obtained with the HFB mean-field states,
and the agreement with the experimental data is better.
Again, the results of projected configuration-mixing and
of the cranked VANP approximation (the solid line)
calculated by Eq.~(\ref{eq:J2mfnp}) are very similar.
\begin{figure}[!htb]
\begin{center}
\includegraphics[width=75mm]{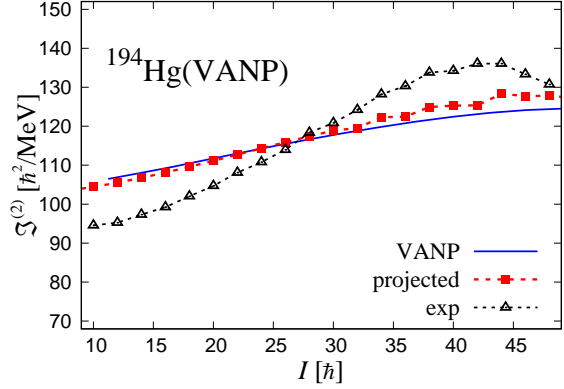}
\end{center}
\vspace*{-8mm}
\caption{(Color online)
Similar to Fig.~\ref{fig:hgmoi} but the mean-field states
obtained by the VANP method are utilized
and the particle-number projection is also performed, cf. Eq.~(\ref{eq:addnp}).
Here the solid line is calculated by Eq.~(\ref{eq:J2mfnp}).
}
\label{fig:hgnpmoi}
\end{figure}

Figure~\ref{fig:hgnpmoimix} shows four $\Jtwom$ moments of inertia
calculated from the projected spectra obtained by
a single cranked VANP mean-field state at each rotational frequency
in addition to the result of final configuration-mixing.
The $\Jtwom$ moments of inertia calculated by the projection
from a single mean-field state with the VANP method are not so different
from those with the HFB method shown in Fig.~\ref{fig:hgmoimix},
although the result of configuration-mixing more rapidly increases
as a function of angular momentum;
this also suggests the importance of multicranked configuration-mixing.
\begin{figure}[!htb]
\begin{center}
\includegraphics[width=75mm]{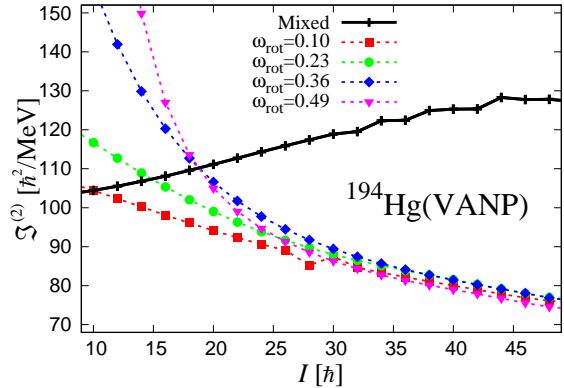}
\end{center}
\vspace*{-8mm}
\caption{(Color online)
Similar to Fig.~\ref{fig:hgmoimix} but the mean-field states
obtained by the VANP method are utilized
and the particle-number projection is also performed, cf. Eq.~(\ref{eq:addnp}).
}
\label{fig:hgnpmoimix}
\end{figure}

To see the effect of configuration-mixing,
we show in Fig.~\ref{fig:hgnpenergy} the four calculated spectra
from a single VANP mean-field state at each rotational frequency
in addition to the result of configuration-mixing.
The absolute energy of the projected configuration-mixing spectrum
from the VANP states is about 3.4 MeV smaller at $I\approx 0$
due to the particle-number projection
on top of the angular-momentum projection.
Again, the result of configuration-mixing follows the envelope curve
of a family of four spectral curves obtained by the projection
from a single cranked VANP mean-field states.
Compared to the results with the HFB mean-field states,
the spin and the energy values at their contacting points are
smaller and larger, respectively, for the VANP method,
and consequently the curvature of the parabolic spectrum of
the final configuration-mixing is larger at lower spin,
while it is smaller at higher spin.
This leads to reduction of the $\Jtwom$ moment of inertia at lower spin
and increase at higher spin
as a result of multicranked configuration-mixing
using the VANP mean-field states.
\begin{figure}[!htb]
\begin{center}
\includegraphics[width=75mm]{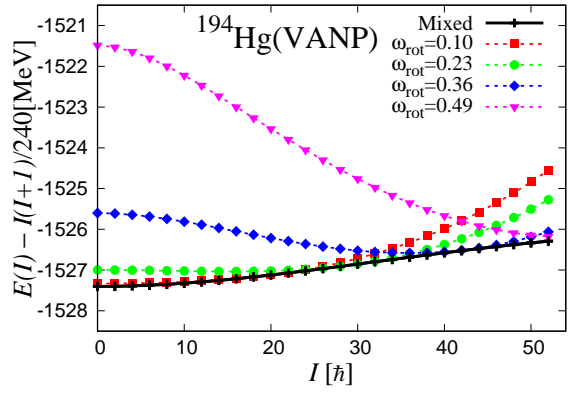}
\end{center}
\vspace*{-8mm}
\caption{(Color online)
Similar to Fig.~\ref{fig:hgenergy} but the mean-field states
obtained by the VANP method are utilized
and the particle-number projection is also performed, cf. Eq.~(\ref{eq:addnp}).
}
\label{fig:hgnpenergy}
\end{figure}
\begin{figure}[!htb]
\begin{center}
\includegraphics[width=75mm]{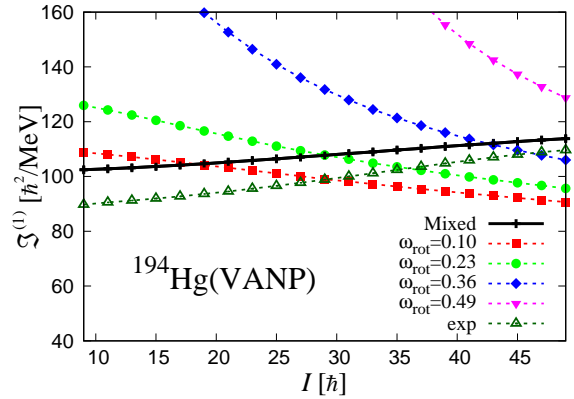}
\end{center}
\vspace*{-8mm}
\caption{(Color online)
Similar to Fig.~\ref{fig:hgj1} but the mean-field states
obtained by the VANP method are utilized
and the particle-number projection is also performed, cf. Eq.~(\ref{eq:addnp}).
}
\label{fig:hgnpj1}
\end{figure}

In the same way, the $\Jonem$ moments of inertia calculated from the spectra
obtained by a single VANP mean-field state at each cranking frequency
and the result of configuration-mixing are displayed in Fig.~\ref{fig:hgnpj1}
corresponding to the spectra in Fig.~\ref{fig:hgnpenergy}.
The experimental $\Jonem$ moment of inertia is also included.
It is clearly seen that the resultant $\Jonem$ moment of inertia
by configuration-mixing employing the VANP mean-field states
is reduced compared with
the result using the HFB mean-field states shown in Fig.~\ref{fig:hgj1}.
Consequently, the agreement with the experimental data
is better in the result with the VANP mean-field states.
This reduction of the $\Jonem$ moment of inertia
is in agreement with the general analysis of the pairing fluctuations
at high-spin states in Refs.~\cite{pairRMP,YRS90},
where the systematic dealignment effect has been recognized.
Thus the effect of dynamic pairing correlations is not
very large also for $^{194}$Hg;
these results are slightly different from those in Ref.~\cite{VER00}.

\subsubsection{No number projection with cranked VANP method}
\label{sec:Hg194VANPnoNP}

The numerical cost to perform both the particle-number and
angular-momentum projection at the same time is very large.
On the other hand, the effect of the number projection is
usually not very large, see e.g. Ref.~\cite{HS95}.
We have also confirmed it in the calculation of the spectrum
for a tetrahedrally deformed nucleus~\cite{TSD13}.
Therefore, we try the multicranked configuration-mixing calculation
with {\it no particle-number projection}
by employing the VANP mean-field states.
The result for the $\Jtwom$ moment of inertia is depicted
in Fig.~\ref{fig:hgwonpmoimix},
where the four $\Jtwom$ moments of inertia
calculated by the projection from a single VANP mean-field state
at each rotational frequency are also included.
The norm cut-off factor $10^{-12}$ can be used in this calculation.
\begin{figure}[!htb]
\begin{center}
\includegraphics[width=75mm]{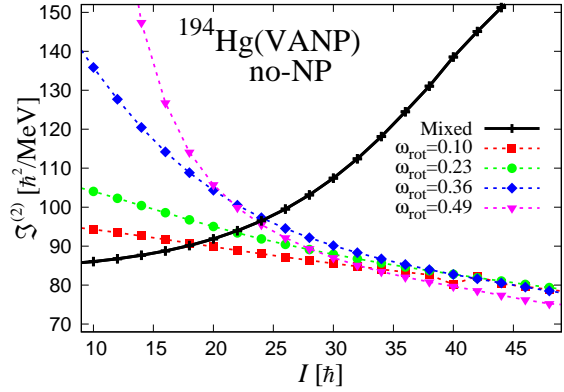}
\end{center}
\vspace*{-8mm}
\caption{(Color online)
Similar to Fig.~\ref{fig:hgnpmoimix} but
no particle-number projection is performed.
}
\label{fig:hgwonpmoimix}
\end{figure}

Compared with the corresponding result shown in Fig.~\ref{fig:hgnpmoimix},
the resultant $\Jtwom$ moment of inertia of configuration-mixing
is very different.
The increase as a function of the rotational frequency is much larger;
even larger than that of the experimental data.
In contrast, those calculated with a single mean-field state
are not very different from the case with particle-number projection.
In order to understand the reason for it,
the four spectra calculated from the projection from a single mean-field state
are displayed in Fig.~\ref{fig:hgwonpenergy}
in addition to the resultant spectra of configuration-mixing.
Note that the absolute energies of the projected spectra
are even larger than those using the HFB states
shown in Fig.~\ref{fig:hgenergy},
because the particle-number projection is not performed
with the VANP mean-field states.
It can be seen that the spectral curves obtained by the projection
from a single mean-field state are not very different from those
calculated with number projection shown in Fig.~\ref{fig:hgnpenergy},
which is consistent with the observation that
the effect of number projection is not very important in Ref.~\cite{TSD13},
where the projection was performed from a single mean-filed state.
However, the resultant spectrum of configuration-mixing
is dramatically changed if no particle-number projection is performed.
\begin{figure}[!htb]
\begin{center}
\includegraphics[width=75mm]{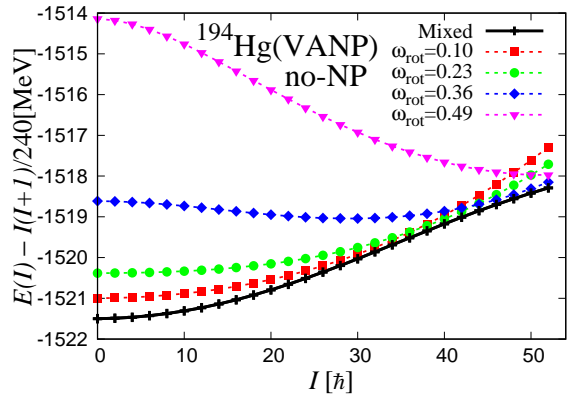}
\end{center}
\vspace*{-8mm}
\caption{(Color online)
Similar to Fig.~\ref{fig:hgnpenergy} but
no particle-number projection is performed.
}
\label{fig:hgwonpenergy}
\end{figure}

It is worthwhile mentioning that the energy gain caused by
the configuration-mixing seen in Fig.~\ref{fig:hgwonpenergy}
is considerably larger than those
in Figs.~\ref{fig:hgenergy} and~\ref{fig:hgnpenergy},
especially at low spin, $I \ltsim 20$.
This indicates that the coupling matrix elements
in the Hill-Wheeler equation in Eq.~(\ref{eq:HW})
are overestimated when the number projection
is not carried out for the VANP mean-field states.
If the HFB mean-field state with cranking frequency $\omega^{(n')}_{\rm rot}$
is represented by the quasiparticle states of
the other HFB state with $\omega^{(n)}_{\rm rot}$,
their coupling matrix elements consists of
the terms between the zero quasiparticle state
and the two, four, six, $\cdots$ quasiparticle states.
However, those between the zero and two quasiparticle states,
which are considered to contribute most
owing to the small energy denominators,
vanish because of the selfconsistency condition of the HFB,
i.e. vanishing ``dangerous terms'', see e.g. Ref.~\cite{RS80}.
A similar result applies true also for the VANP mean-field states but
with respect to ``number projected'' quasiparticle states.
Therefore, the coupling terms of configuration-mixing are supposed to be small
if the mean-field states are determined selfconsistently
as in the cases shown in Fig.~\ref{fig:hgenergy} for the HFB method
and in Fig.~\ref{fig:hgnpenergy} for the VANP method.
This is not the case, however, if the number projection
is neglected for the configuration-mixing using the VANP mean-field states
in Fig.~\ref{fig:hgwonpenergy}.
In this way, although the effect of number projection is not very important
for the projection from a single VANP mean-field state,
its effect can be rather large for the configuration-mixing
like in the present case.
Thus, we should be careful about employing any kind of approximations
which break the selfconsistently of mean-field states
if the multicranked configuration-mixing is performed.
Although the effect of breaking the selfconsistency is found to be
not so large for the case of another nucleus $^{152}$Dy (not shown)
as in the case of $^{194}$Hg,
the same caution should be applied.
In fact, the results of final configuration-mixing do not agree with
the cranked VANP approximation in Eq.~(\ref{eq:J2mfnp})
in both $^{152}$Dy and $^{194}$Hg cases,
if the number projection is not performed for the VANP states.

\section{Conclusion}
\label{sec:concls}

The nuclear mean-field theory is one of the most successful theories
to describe nuclear properties from the microscopic view point.
The key concept is the spontaneous symmetry breaking,
where complicated correlations between the constituent nucleons
can be incorporated through the nuclear mean-field.
To obtain the quantum mechanical eigenstates, however,
one has to restore the symmetry broken in the mean-field:
The nuclear mean-field is the intrinsic state,
from which a series of symmetry-preserving eigenstates
is generated by the quantum-number projection method.
The collective rotation is a good example;
a sequence of eigenstates composing the rotational band is
obtained by the angular-momentum projection
from a single deformed mean-field state.
Although it is conceptually correct and appealing,
the result of projection from a single mean-field state
is not enough for precise description of the rotational band
especially at high-spin states,
which is clearly indicated in the present work.
It is necessary to make multicranked configuration-mixing, i.e.,
several mean-field states with different cranking frequencies
should be properly superposed after angular-momentum projection.

Thus, we show how the approach of projected multicranked configuration-mixing
works for the good description of high-spin rotational bands.
This approach, cf. Eq.~(\ref{eq:PTanz}) or Eq.~(\ref{eq:proj}),
has been originally proposed
by Peierls and Thouless~\cite{PT62} and developed recently
in our previous works~\cite{STS15,STS16}.
In the present work, it is applied to the investigation
of superdeformed bands in the $^{152}$Dy and $^{194}$Hg nuclei,
where long rotational sequences are observed
and the spin-assignments have been provided.
These two representative nuclei are chosen to investigate
the effect of pairing fluctuations on
the $\Jtwom$ moment of inertia of the superdeformed band,
for which large difference has been known
between the $A\approx 150$ and $A\approx 190$ mass regions.
The two methods to incorporate the pairing correlations,
i.e., the HFB and the VANP methods,
are employed to determine the mean-field states,
with which the angular-momentum projection
(and the particle-number projection
at the same time for the VANP method) and
subsequent configuration-mixing is performed.
The Gogny D1S force is used as the effective interaction.
The different behavior of the $\Jtwom$ moment of inertia
in $^{152}$Dy and $^{194}$Hg is attributed to
the effect of pairing correlations,
which is stronger in $^{194}$Hg than in $^{152}$Dy.
This is consistent with other previous works,
see e.g. Ref.~\cite{VER00},
in which the angular-momentum projection is not considered though.

It is demonstrated that the configuration-mixing of
several mean-field states with different cranking frequencies are
essential to understand the $\Jonem$ and $\Jtwom$ moments of inertia
of superdeformed nuclei
by the angular-momentum-projection calculation.
The projection calculation from a single mean-field state obtained
by either the HFB or VANP method does not give any reasonable results;
the calculated $\Jtwom$ moments of inertia are too small and
decrease gradually as functions of angular momentum.
With configuration-mixing after projection,
fair agreement with experimental data is achieved
for both the $\Jonem$ and $\Jtwom$ moments of inertia,
although the agreement is not perfect in the present investigation.
A selfconsistent treatment is emphasized for the configuration-mixing;
namely, both the particle-number and angular-momentum projection
are necessary if the VANP mean-field states are employed.
With proper selfconsistency, however, the results of the multicranked
configuration-mixing for $\Jtwom$ moments of inertia are found
to be very similar to those calculated with the semiclassical
cranked HFB or VANP approximation without angular-momentum projection.
This means that the mean-field approximation (or its extension like VANP)
gives a fairly good approximation for the description of
superdeformed high-spin rotational bands.



\vspace*{10mm}


\end{document}